\documentclass{article}%
\usepackage[top=3cm, bottom=3cm, left=3cm, right=3cm]{geometry}
\usepackage{amsmath}
\usepackage{amsfonts}
\usepackage{amssymb}
\usepackage{graphicx}
\usepackage{cite}%
\providecommand{\U}[1]{\protect\rule{.1in}{.1in}}
\begin{document}

\title{Multichannel decay law}
\author{Francesco Giacosa\\\textit{Institute of Physics, Jan-Kochanowski University, }\\\textit{ul. Uniwersytecka 7, 25-406, Kielce, Poland.}\\\textit{Institute for Theoretical Physics, J. W. Goethe University, }\\\textit{ Max-von-Laue-Str. 1, 60438 Frankfurt, Germany.}}
\date{}
\maketitle

\begin{abstract}
It is well known, both theoretically and experimentally, that the survival
probability for an unstable quantum state, formed at $t=0,$ is not a simple
exponential function, even if the latter is a good approximation for
intermediate times. Typically, unstable quantum states/particles can decay in
more than a single decay channel. In this work, the general expression for the
probability that an unstable state decays into a certain $i$-th channel
between the initial time $t=0$ and an arbitrary $t>0$ is provided, both for
nonrelativistic quantum states and for relativistic particles. These partial
decay probabilities are also not exponential and their ratio turns out to be
not a simple constant, as it would be in the exponential limit. Quite
remarkably, these deviations may last relatively long, thus making them
potentially interesting in applications. Thus, multichannel decays represent a
promising and yet unexplored framework to search for deviations from the
exponential decay law in quantum mechanical systems, such as quantum
tunnelling, and in the context of particle decays.

\end{abstract}

Decay processes are ubiquitous in the realm of Quantum Mechanics (QM) and
Quantum Field Theory (QFT). Quite interestingly, they comprise utterly
different time scales, which range from the very short lifetime of the $\rho$
meson ($\tau\sim10^{-23}$ s) and the Higgs boson ($\tau\sim10^{-22}$ s ), to
the extremely long lifetimes of some nuclei, as for instance the
double-$\beta$ decays of $^{130}$Te with $\tau\sim10^{20}$ y and $^{124}$Xe
with $\tau\sim10^{22}$ y \cite{pdg}. It is then remarkable that `in nuce' a
very similar theory of decay can be applied to such different physical systems.

The survival probability $p(t)$ is the probability that an unstable state in
QM or an unstable particle in QFT created at $t=0$ is still undecayed at the
time $t>0.$ It can be expressed by the following famous formula (e.g. Ref
\cite{fonda} and the following for an explicit QM derivation):%

\begin{equation}
p(t)=\left\vert \int_{E_{th,1}}^{\infty}\mathrm{dE}d_{S}(E)e^{-\frac
{i}{\hslash}Et}\right\vert ^{2}\text{ ,} \label{p}%
\end{equation}
where $d_{S}(E)$ is the the so-called energy (or mass) probability density and
$E_{th,1}$ its minimal low-energy threshold. Namely, an unstable state is not
an energy eigenstate of the underlying Hamiltonian of the system and
$\mathrm{dE}d_{S}(E)$ can be interpreted as the probability that it has an
energy between $E$ and \ $E+\mathrm{dE}$. Note, for a stable state
$d_{S}(E)=\delta(E-M)$: one recovers $p(t)=1$ for each $t.$

The expression in Eq. (\ref{p}) is the starting point of numerous theoretical
works, see e.g. Refs.
\cite{fonda,khalfin,winter,levitan,dicus,peshkin,calderon,koide,fpnew,kk1,kk2,duecan,giacosapra1,raczynska}%
, as well as experimental investigations
\cite{raizen,raizen2,rothe,kelkar,pasclast}. It turns out that the standard
exponential decay law $p(t)\simeq e^{-t/\tau}$ is a very good approximation
for intermediate times, even if this is not an exact law. Here, $\tau$ refers
to the mean lifetime of the unstable state, while $\Gamma=\hslash/\tau$
represents its decay width.

The decay law is typically quadratic for short times, $p(t)\simeq1-t^{2}%
/\tau_{Z}^{2},$ where $\tau_{Z}$ is the so-called Zeno time. As a consequence,
for frequently repeated measurements at very small time intervals, the Zeno
effect, that is the freezing of state in its unstable initial configuration,
takes place \cite{dega,misra,koshino,fp,fptopicalreview,schulman,giacosapra2},
as confirmed in experiments on atomic transitions
\cite{itano,balzer,streed,haroche} as well as on quantum tunnelling
\cite{raizen,raizen2}. For measurements repeated for slightly larger time
intervals, also the inverse Zeno effect, that is an increased decay rate by
measurements, is possible \cite{kk1,kk2,nakazato,gpn}, as verified in\ Ref.
\cite{raizen2}. Finally, at large times a power law sets in, $p(t)\simeq
t^{-\alpha}$ where $\alpha>0$ depends on the details of the interaction that
governs that particular decay \cite{rothe}. All these well-known features deal
with the survival probability of the unstable state expressed in Eq. (\ref{p})
and apply to basically each unstable state.

Next, a simple question might be asked: Which is the probability that the
decay of the unstable state occurs between $t=0$ and the time $t$? The answer
is trivial, since the probability that the decay has actually occurred,
denoted as $w(t)$, must be
\begin{equation}
w(t)=1-p(t)\text{ .}%
\end{equation}
Similarly, the quantity $h(t)=w^{\prime}(t)=-p^{\prime}(t)$ is the probability
decay density, with $h(t)dt$ being the probability that the decay occurs
between $t$ and $t+dt.$

Yet, in many practical cases that apply to both unstable quantum states and
particles, there is not a single decay channel, but $N$ distinct ones.
Typically, excited atoms can decay into different low-lying energy levels and
elementary particles, such as the $\tau$ lepton or the Higgs boson as well as
composite hadrons, display various decay channels. In these cases, in the
exponential (or Breit-Wigner (BW)) limit the decay width $\Gamma$ is the sum
of N distinct terms, $\Gamma=\sum_{i=1}^{N}\Gamma_{i}$, where $\Gamma_{i}$ is
the partial decay width in the $i$-th channel and $\Gamma_{i}/\Gamma$ the
corresponding branching ratio.

Then, a natural, but less easy question is the following: \textit{How to
calculate, in a general fashion, the probability, denoted as }$w_{i}%
(t),$\textit{\ that the decay occurs in the }$i$\textit{-th channel between
}$0$\textit{\ and }$t$?

In the BW limit, the expected result is simply $w_{i}(t)=\frac{\Gamma_{i}%
}{\Gamma}w(t),$ thus each decay probability $w_{i}(t)$ is a fraction of the
total decay probability $w(t)$, but, as we shall see, this is not valid in
general. In the precursory work of Ref. \cite{duecan} a partial approximate
solution for the function $w_{i}(t)$ was put forward in specific models and in
the recent Ref. \cite{2delta} the two-channel decay was studied via tunnelling
to the `left' and to the `right' in an asymmetric double-delta potential. The
aim of this work is to present the general form of the $i$-th channel decay
probability $w_{i}(t)$.

\bigskip

Let us start with the QM case. We consider a system that contains an unstable
state $\left\vert S\right\rangle $ that can decay into final states of the
type $\left\vert E,i\right\rangle ,$where $i=1,...,N$ enumerates the decay
channels ($E$ being the energy of the final state, see below). By denoting
with $H$ the Hamiltonian of the system, the time-evolution operator
$e^{-iHt/\hbar}$ is clearly of crucial importance for our purposes, since it
controls how the unstable state $\left\vert S\right\rangle $ evolves in time.
First, we rewrite it as ($t>0$):%
\begin{equation}
e^{-\frac{i}{\hslash}Ht}=\frac{i}{2\pi}\int_{-\infty}^{+\infty}\mathrm{dE}%
\frac{e^{-\frac{i}{\hslash}Et}}{E-H+i\varepsilon}\text{ ,}%
\end{equation}
out of which the survival probability amplitude is obtained by the expectation
value
\begin{equation}
a(t)=\left\langle S\right\vert e^{-\frac{i}{\hslash}Ht}\left\vert
S\right\rangle =\frac{i}{2\pi}\int_{-\infty}^{+\infty}\mathrm{dE}%
G_{S}(E)e^{-\frac{i}{\hslash}Et}=\int_{E_{1,th}}^{+\infty}\mathrm{dE}%
d_{S}(E)e^{-\frac{i}{\hslash}Et}.\label{a}%
\end{equation}
The survival probability of Eq. (\ref{p}) emerges as $p(t)=\left\vert
a(t)\right\vert ^{2}$. The (formally exact) propagator $G_{S}(E)$ of the
unstable state $\left\vert S\right\rangle $ with energy $M$ (or mass $m,$ with
$M=mc^{2})$ reads%

\begin{equation}
G_{S}(E)=\left\langle S\right\vert \frac{1}{E-H+i\varepsilon}\left\vert
S\right\rangle =\frac{1}{E-M+\Pi(E)+i\varepsilon}=\int_{E_{1,th}}^{+\infty
}\mathrm{dE}^{\prime}\frac{d_{S}(E^{\prime})}{E-E^{\prime}+i\varepsilon}\text{
.} \label{qmprop}%
\end{equation}
The self-energy function $\Pi(E)$ is the one-state irreducible contribution.
Intuitively, it represents the transition $\left\vert S\right\rangle
\rightarrow\left\vert E,i\right\rangle \rightarrow$ $\left\vert S\right\rangle
$ where an integral/sum over $E$ and $i$ is taken. (With no loss of
generality, we adopt the choice $\operatorname{Re}\Pi(M)=0$ by a suitable
subtraction, thus $M$ is the nominal energy of the state or, more formally,
the value at which $\operatorname{Re}[G_{S}^{-1}(E=M)]=0$). As a consequence
of the optical theorem (e.g. Ref. \cite{peskin}), the decay width function
$\Gamma(E)=2\operatorname{Im}\Pi(E)$ depends on the energy (only in the
exponential limit it is a simple constant). The `on shell' BW decay width is
recovered as an approximation by setting $E=M$, thus $\Gamma=\Gamma(M)$.

The spectral function (or mass distribution) $d_{S}(E)$, already introduced in
the very first Eq. (\ref{p}), is obtained from the last equality of Eq.
(\ref{qmprop}) as:%

\begin{equation}
d_{S}(E)=-\frac{1}{\pi}\operatorname{Im}G_{S}(E)=\frac{\Gamma(E)}{2\pi
}\left\vert G_{S}(E)\right\vert ^{2}.
\end{equation}
Intuitively, Eq. (\ref{qmprop}) shows that the propagator $G_{S}(E)$ is
rewritten as the `sum' of free propagators with `weight' $d_{S}(E^{\prime
})\mathrm{dE}^{\prime}$ , which as mentioned above is naturally interpreted as
the probability that the state $\left\vert S\right\rangle $ has an energy
contained in the range $(E,E+dE)$. As a consequence, the probability decay
amplitude in Eq. (\ref{a}) is just the integral over $d_{S}(E)$ weighted by
the corresponding time-evolution factor $e^{-iEt/\hslash}$. The normalization
$\int_{E_{th,1}}^{\infty}\mathrm{dE}d_{S}(E)=1$ is guaranteed for any physical
problem (for an explicit proof, see e.g. Refs. \cite{leerev,lupofermionico}).

When $N$ decay channels are available, $\Pi(E)$ is given by the sum over
them:
\begin{equation}
\Pi(E)=\sum_{i=1}^{N}\Pi_{i}(E)\text{ , }\Gamma_{i}(E)=2\operatorname{Im}%
\Pi_{i}(E)\text{ ,}%
\end{equation}
where $\Pi_{i}(E)$ encodes the loop $\left\vert S\right\rangle \rightarrow
\left\vert E,i\right\rangle \rightarrow$ $\left\vert S\right\rangle $ for a
fixed decay channel and $\Gamma_{i}(E)$ is the $i$-th decay width function,
that vanishes below the corresponding threshold $E_{th,i}.$ Here, we assume
for definiteness that $E_{th,1}\leq E_{th,2}\leq...\leq E_{th,N}.$ The
functions $\Gamma_{i}(E)$ are -in principle- obtainable for a given problem,
once the particular Hamiltonian is known. Of course, in practice they are
often not known exactly and their explicit determination may be confined to a
certain energy range and within the validity of the used approach to evaluate
them (as, for example, perturbation theory). The real part $\operatorname{Re}%
\Pi_{i}(E)$ can be obtained by dispersion relations\footnote{By using
dispersion relation, for $E$ complex $\Pi_{i}(E)=-\frac{1}{\pi}\int_{E_{th,i}%
}^{+\infty}\mathrm{dE}^{\prime}\frac{\operatorname{Im}\Pi_{i}(E^{\prime}%
)}{E-E^{\prime}+i\varepsilon}+C,$where $C$ is a subtraction constant.
$\operatorname{Re}\Pi_{i}(E)$ is obtained by taking the principal part. In
some models, it may be convenient to calculate directly the loop function
$\Pi_{i}(E)$.}. The partial BW widths are $\Gamma_{i}=\Gamma_{i}(M)$ (with
$\Gamma=\Gamma(M)=\sum_{i=1}^{N}\Gamma_{i}$ being the total width
$\tau=\hslash/\Gamma$ the mean lifetime).

We also rewrite the spectral function as the sum over partial spectral functions%

\begin{equation}
d_{S}(E)=\sum_{i=1}^{N}d_{S}^{(i)}(E)\text{ with }d_{S}^{(i)}(E)=\frac
{\Gamma_{i}(E)}{2\pi}\left\vert G_{S}(E)\right\vert ^{2}\text{,}%
\end{equation}
where $\mathrm{dE}d_{S}^{(i)}(E)$ is the probability that the state
$\left\vert S\right\rangle $ has an energy between $(E,E+dE)$ \textit{and}
decays in the $i$-th channel. Then, the integral
\begin{equation}
\int_{E_{th,i}}^{\infty}\mathrm{dE}d_{S}^{(i)}(E)=r_{i} \label{ri}%
\end{equation}
is easily interpreted as the asymptotic branching ratio for the decay in the
$i$-th channel. In models, the BW branching ratio $\Gamma_{i}/\Gamma$
represents a good (although not exact) estimate of the value $r_{i},$ see
later on for an example.

Once the functions $\Gamma_{i}(E)$ are \textit{(assumed to be)} known, we can
map the quantum decay of the unstable state $\left\vert S\right\rangle $ onto
an effective Lee Hamiltonian, see e.g.
\cite{duecan,leerev,lee,lee2,jc,qcdeff,scully,sherman,xiaozhou,xiaozhou2,pok},
which reproduces the propagator of Eq. (\ref{qmprop}) exactly:
\begin{equation}
H_{\text{L}}=M\left\vert S\right\rangle \left\langle S\right\vert \text{
}+\sum_{i=1}^{N}\int_{E_{i,th}}^{\infty}\mathrm{dE}E\left\vert
E,i\right\rangle \left\langle E,i\right\vert +\sum_{i=1}^{N}\int_{E_{i,th}%
}^{\infty}\mathrm{dE}\sqrt{\frac{\Gamma_{i}(E)}{2\pi}}\left(  \left\vert
E,i\right\rangle \left\langle S\right\vert +h.c.\right)  \text{.} \label{lee}%
\end{equation}
The ket $\left\vert E,i\right\rangle $ is the state describing the $i$-th
channel decay product with $E$ being the eigenvalue of the non-interacting
part of $H_{\text{L}}$. Clearly, the difficulty lies in the previous
determination of the decay functions $\Gamma_{i}(E)$, which are used here as
an input. The searched probability $w_{i}(t)$ that the unstable state decays
in the $i$-th channel between $0$ and $t$ is formally defined as:
\begin{equation}
w_{i}(t)=\int_{E_{th,i}}^{\infty}\mathrm{dE}\left\vert \left\langle
E,i\left\vert e^{-\frac{i}{\hslash}H_{\text{L}}t}\right\vert S\right\rangle
\right\vert ^{2}\text{ .}%
\end{equation}
Namely, it is the probability that the original state $\left\vert
S\right\rangle $ has evolved into the decay product $\left\vert
E,i\right\rangle $ at the time $t>0$. An explicit calculation of the matrix
element in the integrand delivers \cite{duecan,leerev}%

\begin{equation}
\left\langle E,i\left\vert e^{-iH_{\text{L}}t}\right\vert S\right\rangle
=\sqrt{\frac{\Gamma_{i}(E)}{2\pi}}\frac{i}{2\pi}\int_{-\infty}^{+\infty
}\mathrm{dE}^{\prime}\frac{e^{-\frac{i}{\hslash}E^{\prime}t}}{E-E^{\prime
}+i\varepsilon}G_{S}(E^{\prime})\text{ ,}%
\end{equation}
thus $w_{i}(t)$ is obtained as:
\begin{equation}
w_{i}(t)=\int_{E_{th,i}}^{\infty}\mathrm{dE}\frac{\Gamma_{i}(E)}{2\pi
}\left\vert \frac{i}{2\pi}\int_{-\infty}^{+\infty}\mathrm{dE}^{\prime}%
\frac{G_{S}(E^{\prime})e^{-\frac{i}{\hslash}E^{\prime}t}}{E-E^{\prime
}+i\varepsilon}\right\vert ^{2}\text{.} \label{wi0}%
\end{equation}
Upon introducing the spectral function as in\ Eq. (\ref{qmprop}), $w_{i}(t)$
can be expressed as:%

\begin{equation}
w_{i}(t)=\int_{E_{th,i}}^{\infty}\mathrm{dE}\frac{\Gamma_{i}(E)}{2\pi
}\left\vert \int_{E_{th,1}}^{+\infty}\mathrm{dE}^{\prime}d_{S}(E^{\prime
})\frac{e^{-\frac{i}{\hslash}E^{\prime}t}-e^{-\frac{i}{\hslash}Et}}{E^{\prime
}-E}\right\vert ^{2}\text{ ,} \label{wi1}%
\end{equation}
or equivalently%
\begin{equation}
w_{i}(t)=\int_{E_{th,i}}^{\infty}\mathrm{dE}\frac{\Gamma_{i}(E)}{2\pi
}\left\vert \int_{0}^{t}d\tau\frac{a(\tau)}{\hslash}e^{\frac{i}{\hslash}E\tau
}\right\vert ^{2}\text{ .} \label{wi2}%
\end{equation}
The expressions $w_{i}(t)$ in Eqs. (\ref{wi0}), (\ref{wi1}), (\ref{wi2})
represent\textit{\ the main result }of this work and can be easily evaluated
numerically, since they involve only calculable quantities, such as the
$\Gamma_{i}(E)$ and $M$ (which fix also $d_{S}(E)$ and $a(t)$). Moreover, the
limiting case $t\rightarrow\infty$ can be calculated by taking into account
that $\left\vert \int_{0}^{\infty}\frac{d\tau}{\hslash}a(\tau)e^{\frac
{i}{\hslash}E\tau}\right\vert =\left\vert G_{S}(E)\right\vert $:
\begin{equation}
\text{ }w_{i}(t\rightarrow\infty)=\int_{E_{th,i}}^{\infty}\mathrm{dE}%
\frac{\Gamma_{i}(E)}{2\pi}\left\vert G_{S}(E)\right\vert ^{2}=\int_{E_{th,i}%
}^{\infty}\mathrm{dE}d_{S}^{(i)}(E)=r_{i}\text{ ,}%
\end{equation}
the latter being the asymptotic branching ratio in the $i$-th channel of Eq.
(\ref{ri}). Moreover, from Eq. (\ref{wi1}) one has $w_{i}(0)=0$ and the small
$t$ expansion reads $w_{i}(t)\simeq c_{i}t^{2}$ with $2\pi\hbar^{2}c_{i}%
=\int_{E_{th,i}}^{\infty}\mathrm{dE}\Gamma_{i}(E)$ (one recovers also the Zeno
time as $2\pi\hbar^{2}\tau_{Z}^{-2}=\int_{E_{th,1}}^{\infty}\mathrm{dE}%
\Gamma(E)$). The ratio $w_{1}/w_{2}\simeq c_{1}/c_{2}$ applies for short
times, but $w_{1}/w_{2}$ is approximated by $\Gamma_{1}/\Gamma_{2}$ at
intermediate times, showing that in general it cannot be a constant, as
confirmed below in a numerical example. Of course, the equality $w=\sum
_{i=1}^{N}w_{i}=1-p(t)$ holds for each $t$. Another quantity of interest is
the partial $i$-th probability decay density $h_{i}(t)$, evaluated as
$h_{i}(t)=w_{i}^{\prime}(t)$: $h_{i}(t)dt$ is the probability that the
unstable quantum state decays in the $i$-th channel between $t$ and $t+dt.$
Clearly, $h(t)=-p^{\prime}(t)=\sum_{i=1}^{N}h_{i}(t)$ also applies.

In the renowned BW limit \cite{ww,ww2,ww3}, one has $\Gamma_{i}(E)=\Gamma_{i}$
(constant) together with $\operatorname{Re}\Pi_{i}(E)=0$, hence
\begin{equation}
d_{S}(E)=\frac{\Gamma}{2\pi}\frac{1}{(E-M)^{2}+\Gamma^{2}/4}\text{ , }%
d_{S}^{(i)}(E)=\frac{\Gamma_{i}}{\Gamma}d_{S}(E) \label{bw1}%
\end{equation}
with $\Gamma=\sum_{i=1}^{N}\Gamma_{i},$ thus $r_{i}=\Gamma_{i}/\Gamma$ is
exact in this limit. The temporal evolution is exactly exponential with
$a(t)=e^{-i\frac{M}{\hslash}t-\frac{\Gamma}{2\hslash}t}$ and $p(t)=e^{-\frac
{\Gamma}{\hslash}t}$ as well as:%
\begin{equation}
w_{i}(t)=\frac{\Gamma_{i}}{\Gamma}w(t)=\frac{\Gamma_{i}}{\Gamma}\left(
1-e^{-\frac{\Gamma}{\hslash}t}\right)  \text{ },\text{ }h_{i}(t)=\frac
{\Gamma_{i}}{\Gamma}h(t)=\frac{\Gamma_{i}}{\Gamma}e^{-\frac{\Gamma}{\hslash}%
t}\text{.} \label{bw2}%
\end{equation}
As already anticipated, the limit $w_{i}(\infty)=r_{i}$ is clearly fulfilled.
In the BW limit, one has:
\begin{equation}
\frac{w_{i}(t)}{w_{j}(t)}=\frac{h_{i}(t)}{h_{j}(t)}=\frac{\Gamma_{i}}%
{\Gamma_{j}}=const. \label{bw3}%
\end{equation}
This is however not true in general, see the following numerical examples and
Refs. \cite{duecan,2delta}.

A useful (although not exact) approximation presented in\ Ref. \cite{duecan}
is given by
\begin{equation}
w_{i}(t)\simeq w_{i}^{\text{appr}}(t)=r_{i}-\operatorname{Re}[a_{i}a^{\ast
}]\text{ },\text{ }a_{i}(t)=\int_{E_{th,i}}^{+\infty}\mathrm{dE}d_{S}%
^{(i)}(E)e^{-\frac{i}{\hslash}Et}\text{ .} \label{wiappr}%
\end{equation}
It fulfills all the required limits ($w_{i}(t)=0,$ $w_{i}(\infty)=r_{i},$
$p+\sum_{i=1}^{N}w_{i}=1,$ as well as the BW one) and, as numerical tests
show, is close to the full result of Eq. (\ref{wi1}). The functions
$w_{i}^{\text{appr}}(t)$ are easier to evaluate because they involve only
one-dimensional integrals and can be useful in first approximation.

\begin{center}
$%
{\parbox[b]{3.3183in}{\begin{center}
\includegraphics[
height=1.8144in,
width=3.3183in
]%
{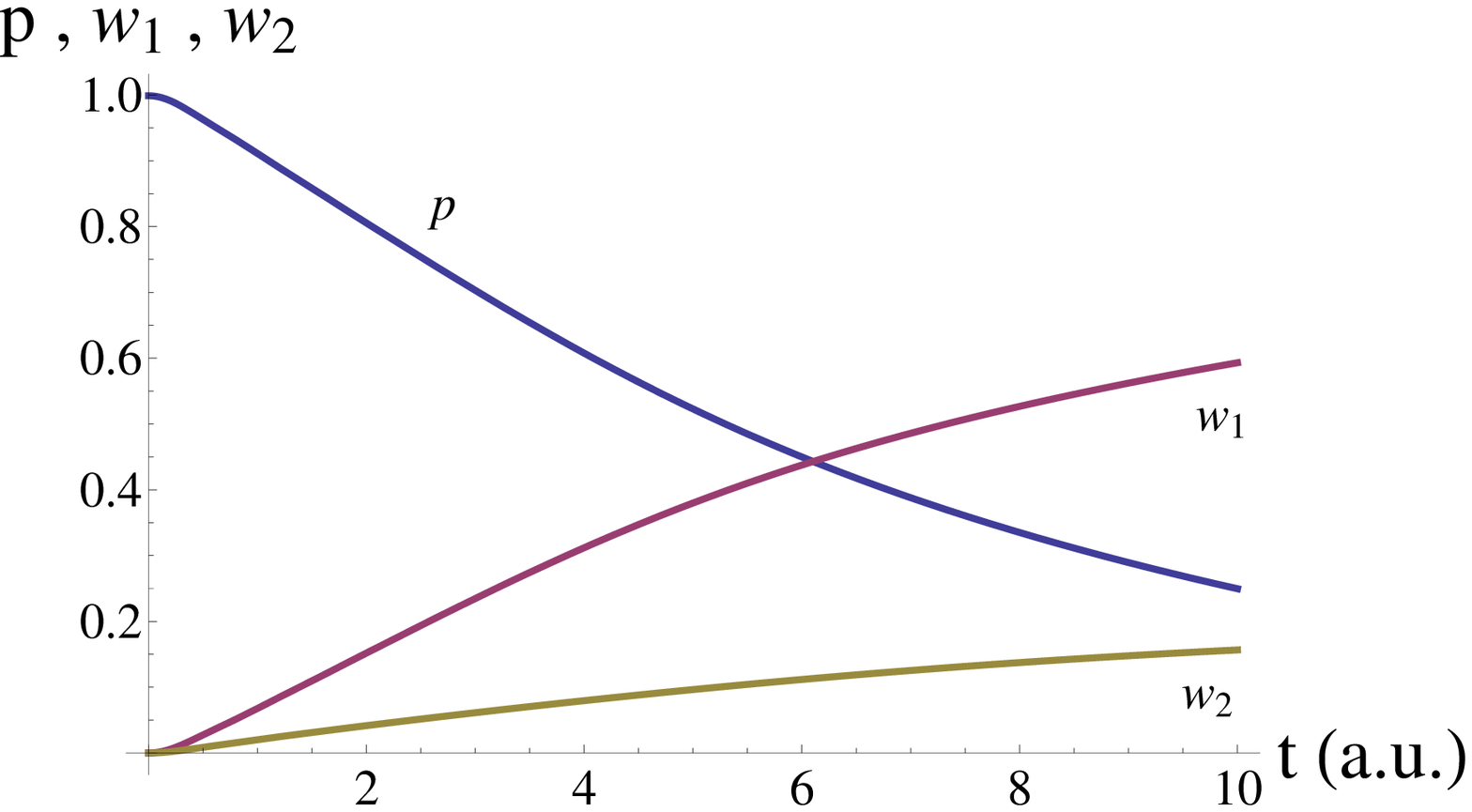}%
\\
The survival probability $p(t)$ of Eq. (\ref{p}) and the decay probabilities
$w_1(t)$ and $w_2(t)$ of Eq. (\ref{wi1}) are plotted as function of $t.$ The
constraint $p+w_1+w_2=1$ holds. Note, $t$ is expressed in a.u. of $[M^-1]$.
\end{center}}}%
$

\bigskip

$%
{\parbox[b]{3.3036in}{\begin{center}
\includegraphics[
height=1.7123in,
width=3.3036in
]%
{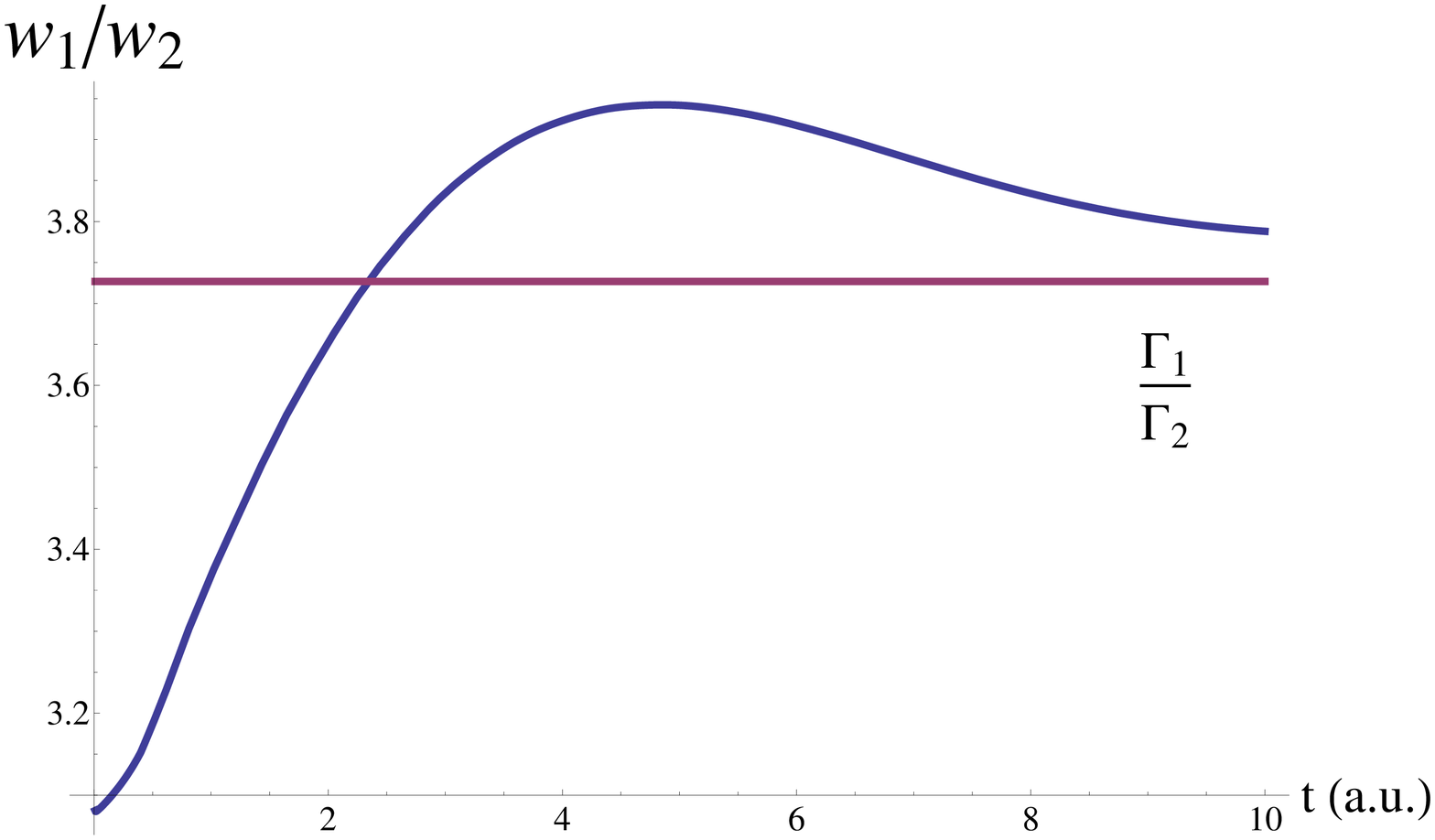}%
\\
Fig. 2: The ratio $w_1/w_2$ is plotted as function of $t.$ The straight line
corresponds to the BW limit $\Gamma_1/\Gamma_2,$ see Eq. (\ref{bw3}).
\end{center}}}%
$
\end{center}

For illustrative purposes, a numerical example that contains two decay
channels is presented in\ Figs. 1-4 for $\Gamma_{i}(E)=2g_{i}^{2}\frac
{\sqrt{E-E_{th,i}}}{E^{2}+\Lambda^{2}}$. This is a rather simple model that
however contains the main needed features that each decay should have, see
e.g. the study of the decay of an excited hydrogen atom in\ Ref.
\cite{pascazioidrogeno}. In particular, the square root $\sqrt{E-E_{th,i}}$
appears quite naturally in various applications as a phase-space term.

By using the numerical values $\hslash=c=1,$ $g_{1}/\sqrt{M}=1,$ $g_{2}%
/\sqrt{M}=0.6,$ $E_{th,1}/M=1/10,$ $E_{th,2}/M=1/2,$ $\Lambda/M=4$, one gets
$r_{1}=w_{1}(\infty)=0.790$ and $r_{2}=w_{2}(\infty)=0.210$, out of which
$r_{1}/r_{2}=3.770$. Interestingly, the corresponding approximate BW results
are very similar: $\Gamma_{1}/\Gamma=0.788\simeq r_{1}$ and $\Gamma_{2}%
/\Gamma=0.212\simeq r_{2},$ with $\Gamma_{1}/\Gamma_{2}=3.727\simeq
r_{1}/r_{2}.$

Fig. 1 shows the (non-exponential) survival probability $p(t)$ evaluated via
the standard Eq. (\ref{p}) as well as $w_{1}(t)$ and $w_{2}(t)$ evaluated
through the novel Eq. (\ref{wi1}). In Fig. 2 the ratio $w_{1}/w_{2}$ shows
sizable deviations from the simple BW constant ratio $\Gamma_{1}/\Gamma_{2}$
of Eq. (\ref{bw3}) (this is so also at intermediate times when $p(t)$ is well
described by an exponential function). Fig. 3 shows the decay probability
densities, $h(t)=-p^{\prime}(t)$ and $h_{i}=w_{i}^{\prime}(t).$ It is evident
that all these functions are much different from the BW form of Eq.
(\ref{bw2}). Moreover $h_{i}(t\rightarrow0)=0$ implies a decreased decay rate
at short time. Finally, Fig. 4 presents the ratio $h_{1}/h_{2},$ which is also
different from the simple BW\ straight line $\Gamma_{1}/\Gamma_{2}$.

\begin{center}
$%
{\parbox[b]{3.5725in}{\begin{center}
\includegraphics[
height=1.6596in,
width=3.5725in
]%
{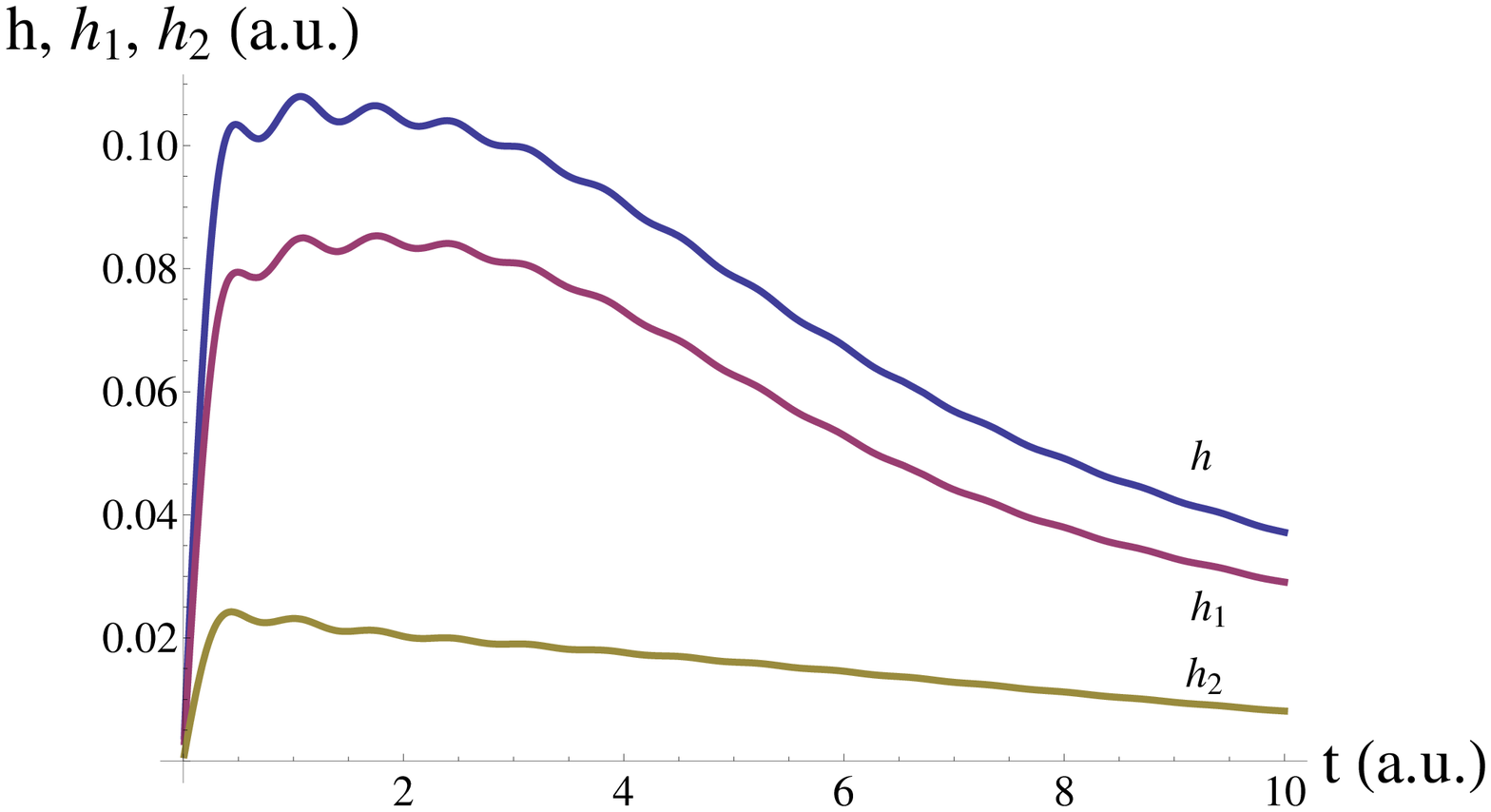}%
\\
Fig. 3: The quantity $h(t)=w^\prime(t)=-p^\prime(t)$ as well as $%
h_i(t)=w_i^\prime(t)$ are plotted. The equality $h(t)=h_1(t)+h_2(t)$ holds.
Note, $h$ and $h_i$ are in units of $[M]$.
\end{center}}}%
$

\bigskip

$%
{\parbox[b]{3.1678in}{\begin{center}
\includegraphics[
height=1.6501in,
width=3.1678in
]%
{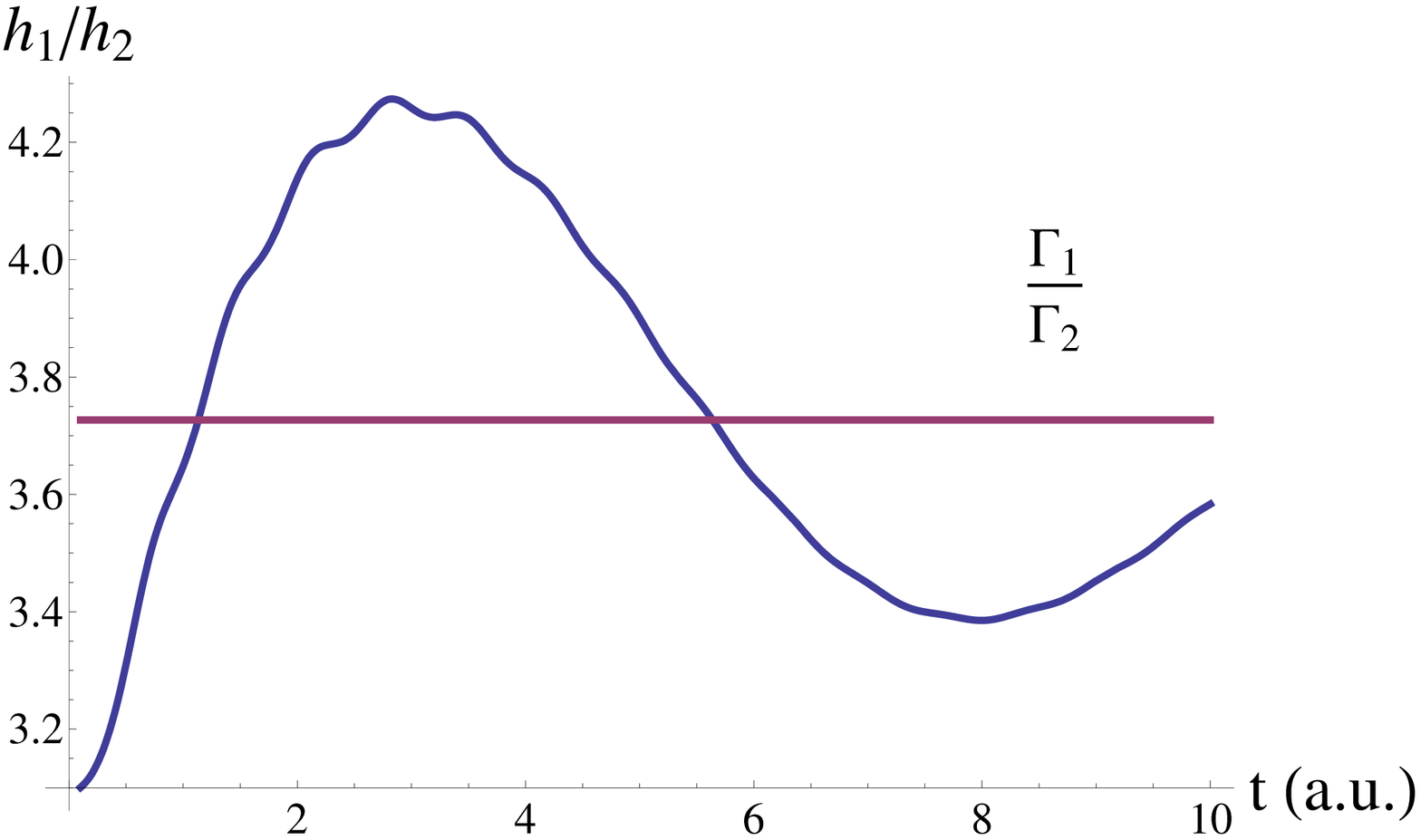}%
\\
Fig. 4: Ratio $h_1/h_2$ as function of $t.$ The straight line corresponds to
the BW limit $\Gamma_1/\Gamma_2$, see Eq. (\ref{bw3}). For the time intervals
where $h_1/h_2>\Gamma_1/\Gamma_2,$ the decay in the first channel is enhanced
(the opposite is true for $h_1/h_2<\Gamma_1/\Gamma_2$).
\end{center}}}%
$
\end{center}

It should be stressed that the precise form of the plotted functions depends
on the chosen physical system; moreover, in the presented example the
functions $\Gamma_{i}(E)$ are assumed to be perfectly known, what in actual
physical systems is generally not the case. Nevertheless, the most relevant
features, such as the departures from the simple BW behavior, with special
attention on the ratios $w_{1}/w_{2}$ and $h_{1}/h_{2}$ depicted in Fig. 2 and
Fig. 4, are expected to be valid in general.

\bigskip

Next, we briefly present the main features that concern the extension to QFT.
In practical terms, when going from QM to QFT, one needs to replace the energy
$E$ with the Mandelstam variable $s=E^{2}$. (Intuitively, in QFT there are
positive and negative energy solutions and the QM free propagator $(E-M)^{-1}$
becomes $(s-M^{2})^{-1}$). The general QFT propagator (in natural units) of an
unstable particle $S$ reads (see e.g. Ref. \cite{peskin}):
\begin{equation}
G_{S}(s)=[s-M^{2}+\Pi(s)+i\varepsilon]^{-1}\text{ , }%
\end{equation}
where $\Pi(s)=\sum_{i=1}^{N}\Pi_{i}(s)$ is the sum of the self energies for
$N$ distinct decay channels. The partial decay function reads
now\footnote{This is also due\textbf{ }to\textbf{ }$E\rightarrow s$\textbf{:
}while the BW pole in QM is $E=M-i\Gamma/2,$ in QFT it is $E^{2}%
=M^{2}-iM\Gamma,$ which reduces to $E\simeq M-i\Gamma/2$ for small $\Gamma$.}
$\Gamma_{i}(s)=\operatorname{Im}\Pi_{i}(s)/\sqrt{s}$. The partial (on shell)
BW decay widths are $\Gamma_{i}=\operatorname{Im}\Pi_{i}(M^{2})/M$ and the
total one $\Gamma=\operatorname{Im}\Pi(M^{2})/M$ .

Just as in the QM case, the spectral function is the imaginary part of the
propagator, $d_{S}(s)=-\frac{1}{\pi}\operatorname{Im}G_{S}(s)$, thus as
function of the energy $d_{S}(E)=-\frac{2E}{\pi}\operatorname{Im}G_{S}%
(s=E^{2})$ \cite{salam,lupo}, where the factor $2E$ arises by the variable
change. The normalization $\int_{E_{i,th}}^{\infty}\mathrm{dE}d_{S}(E)=1$
holds also in\ QFT \cite{lupofermionico} and the survival probability $p(t)$
takes the same form of Eq. (\ref{p}) \cite{nonexpqft,vanhove}. A complete
study of the QFT case with all technical details is left as an important
outlook of the present work.

In QFT, the determination of $\Gamma_{i}(s)$ (and then $\Pi(s)$) is, in
general, a highly nontrivial problem. Perturbative approaches are valid for
small $s,$ but fail for large energies (this is the main reason why
perturbative approach do not capture deviations from the exponential
\cite{maiani}), where nonperturbative methods (such as unitarization schemes
\cite{oller,pelaezrev,hayashi}) are needed. Just as in QM, we assume that a
suitable form of $\Gamma_{i}(s)$ is known for a certain specific problem.
Then, analogously to Eq. (\ref{wi1}), the final result for $w_{i}(t)$ is:%

\begin{equation}
w_{i}(t)=\int_{E_{th,i}}^{\infty}\mathrm{dE}\frac{2E^{2}\Gamma_{i}(E)}{\pi
}\left\vert \int_{E_{th,1}}^{\infty}\mathrm{dE}^{\prime}d_{S}(E^{\prime}%
)\frac{e^{-iE^{\prime}t}-e^{-iEt}}{E^{\prime2}-E^{2}}\right\vert ^{2}\text{ ,}
\label{wiqft}%
\end{equation}
which can be easily evaluated numerically. The approximate expression
$w_{i}^{\text{appr}}(t)=r_{i}-\operatorname{Re}[a_{i}a^{\ast}]$ is also
applicable in QFT. A useful testing case is realized for $\Gamma_{i}%
(s)=g_{i}^{2}\sqrt{(s-s_{th,i})/s}$ with $s_{th,i}=E_{th,i}^{2}$ (the real
part is a constant that can be set to zero \cite{sill}), which gives a similar
qualitative behavior of the curves depicted in Figs. 1-4. Interestingly, this
choice for $\Gamma_{i}(s)$ has been recently successfully applied to the
description of some strongly decaying resonances ($\rho$-meson, $a_{1}$-meson,
$\Delta$-baryon) in Ref. \cite{sill}, delivering better fits than
(non)relativistic BW functions.

\bigskip

In conclusion, for the case of an unstable state with $N$ different available
decay channels, we have provided the expressions for the decay probability in
each channel in QM (Eqs. (\ref{wi1}) and (\ref{wi2})) and QFT (Eq.
(\ref{wiqft})) and we have presented an illustrative numerical example in
Figs. 1-4. These equations can be tested in various experimental setups, such
as an asymmetric tunneling potential, along the line of the already performed
experiment in\ Ref. \cite{raizen} and in agreement with the recent simulation
of Ref. \cite{2delta}. Presently, the possibility to manipulate potentials,
e.g. Refs. \cite{jochim,fallani,kuzmenko,exp1,exp2,exp3,exp4,sowinski}, opens
up interesting developments. Alternatively, the experiment described in\ Ref.
\cite{pasclast} is also very promising: the non-exponential decay is mapped
into arrays of single-mode optical devices, in which the time evolution
corresponds to spatial evolution; it is therefore conceivable to realize a
similar experiment by engineering two decay channels, in such a way to
`measure' the decay probabilities $w_{i}(t).$

The study of multiple decay channels for elementary particles is also an
important and ambitious goal, even though it is typically difficult to measure
deviations from the exponential decay due to the short lifetimes involved. The
ratios $w_{i}/w_{j}$ and $h_{i}/h_{j}$ are expected to deviate longer from the
BW limit and could therefore be the key to see such deviations in such
fundamental systems.

\bigskip

\textbf{Acknowledgments:} the author thanks L. Tinti, G. Pagliara and S.
Mr\'{o}wczy\'{n}ski for stimulating and useful discussions. Financial support
from the OPUS project 2019/33/B/ST2/00613 is acknowledged.

\end{document}